\documentclass[letterpaper,  
              ]{jacow}
\makeatletter%
	\ifboolexpr{bool{xetex}}
	 {\renewcommand{\Gin@extensions}{.pdf,%
	                    .png,.jpg,.bmp,.pict,.tif,.psd,.mac,.sga,.tga,.gif,%
	                    .eps,.ps,%
	                    }}{}
\makeatother

\ifboolexpr{bool{xetex} or bool{luatex}} 
 {}                                      
 {\usepackage[utf8]{inputenc}}           

\usepackage[USenglish]{babel}			 

\usepackage[final]{pdfpages}
\usepackage{multirow}
\usepackage{ragged2e}
\usepackage{lipsum}
\let\OLDthebibliography\thebibliography
\renewcommand\thebibliography[1]{
  \OLDthebibliography{#1}
  \setlength{\parskip}{0pt}
  \setlength{\itemsep}{0pt plus 0.3ex}
}

\ifboolexpr{bool{jacowbiblatex}}%
 {%
  \addbibresource{jacow-test.bib}
  \addbibresource{biblatex-examples.bib}
 }{}
\begin{document}

\title{First Demonstration of Ionization Cooling in MICE}

\author{Tanaz Angelina Mohayai\thanks{tmohayai@hawk.iit.edu}, Illinois Institute of Technology, Chicago IL, U.S. \\  
		on behalf of the MICE Collaboration
		}
	
\maketitle

\begin{abstract}
The Muon Ionization Cooling Experiment (MICE) at Rutherford Appleton Laboratory has studied ionization cooling of muons. Several million individual muon tracks have been recorded passing through a series of focusing magnets and a liquid hydrogen (LH$_{2}$) or lithium hydride (LiH) absorber in a variety of magnetic configurations.~Identification and measurement of muon tracks upstream and downstream of the absorber are used to study the evolution of the $4$D (transverse) emittance. This paper presents and discusses these results.
\end{abstract}
\begin{figure*}[!tbh]
    \centering
    \includegraphics*[width=\textwidth]{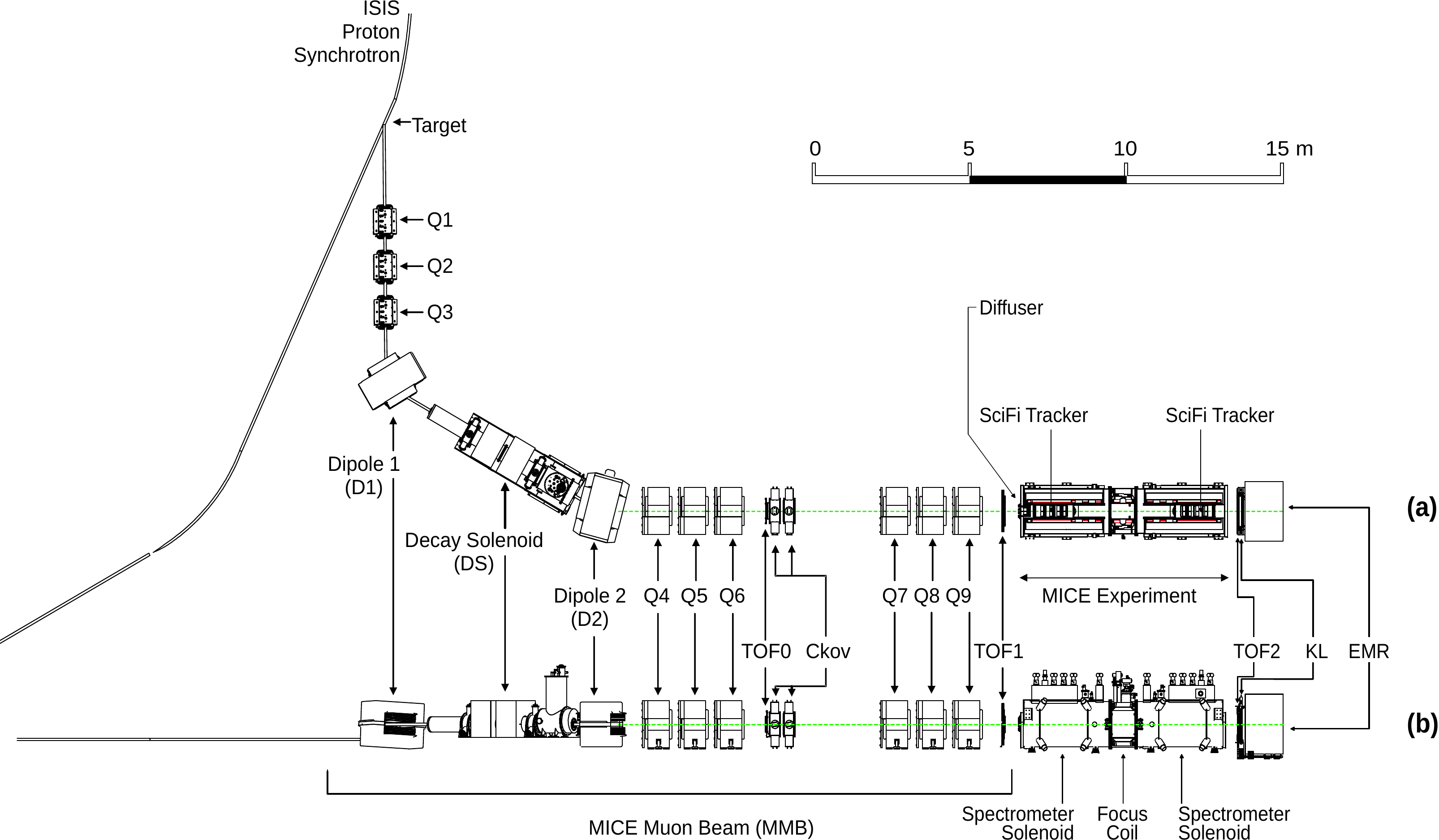}
    \caption{Schematic of the MICE beamline and MICE experiment~\cite{beamline}. (a) and (b) are the top and side views, respectively.}
    \label{fig:beamline}
\end{figure*}
\section{Introduction}
One way of producing muons is to collide a high power proton beam with a target, producing pions which then decay into muons, forming large phase-space volume (emittance) beams. The typical emittance range of a highly diffuse muon beam is between $15$ and $20$ $\pi$ mm$\cdot$rad at production~\cite{note}. In order to optimize muon yield and fit the beam into cost-effective apertures (for a future muon collider or neutrino factory), one would need to reduce the beam emittance. The desired muon normalized beam emittance at the neutrino factory ranges between $2$ and $5$ $\pi$ mm$\cdot$rad~\cite{note}. An initial muon collider needs further cooling with a desired normalized transverse emittance of $0.4$ $\pi$ mm$\cdot$rad and normalized longitudinal emittance of $1$ $\pi$ mm$\cdot$rad \cite{note}. Ionization cooling is the only beam cooling technique suitable for reducing the muon beam emittance within the short muon lifetime~\cite{dave}. This technique works for muons since they interact with matter electromagnetically but without initiating an electromagnetic shower. The international Muon Ionization Cooling Experiment (MICE) is the first experiment demonstrating this technique.
\section{MICE Beamline}
\begin{figure*}[!tbh]
    \centering
    \includegraphics*[width=\textwidth]{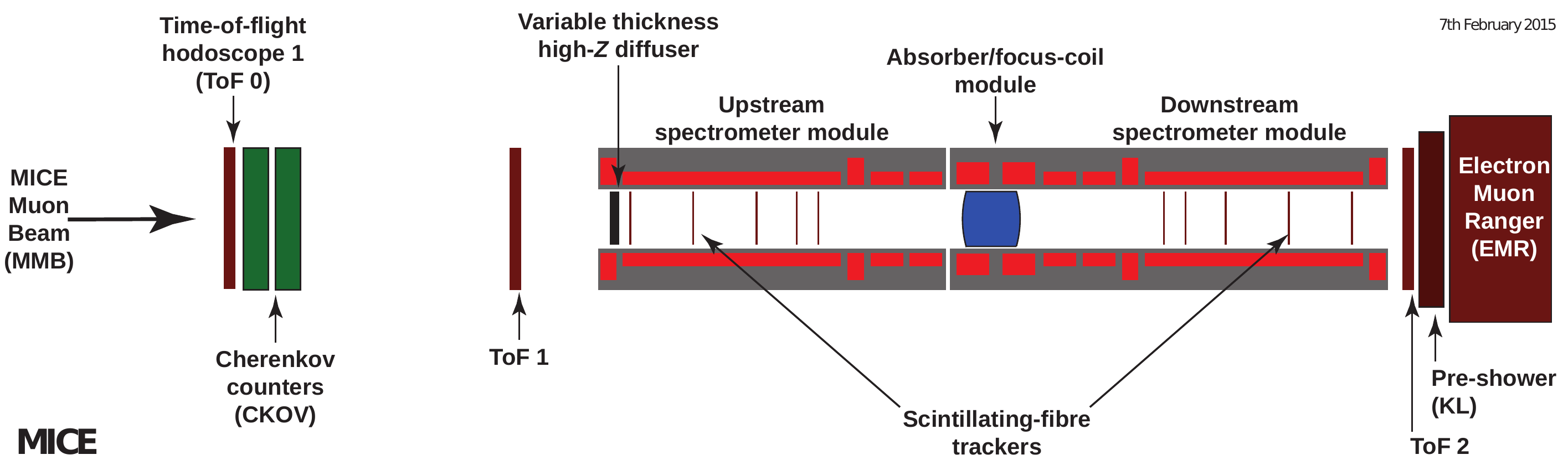}
    \caption{Schematic diagram of MICE in its final configuration.}
    \label{fig:MICE_Step4}
\end{figure*}

The proton beam used for muon production in MICE is  produced by the Rutherford Appleton Laboratory ISIS proton synchrotron in Oxfordshire, U.K. At ISIS, protons are injected into the synchrotron with a kinetic energy of $70$ MeV and are accelerated to $800$ MeV over a period of $10$ ms. MICE operates parasitically to the operation of ISIS and its titanium target interacts with the proton beam~\cite{target}, producing pions and other hadrons. A vacuum window in the ISIS beam pipe allows particles to pass into the MICE beamline (Fig.~\ref{fig:beamline}). In order to avoid losing pions via scraping, a triplet of quadrupole magnets (Q1, Q2, and Q3) are used. Two dipole magnets (D1 and D2), with one located in the ISIS synchrotron vault and the other at the exit of the MICE decay solenoid (DS), are used for pion and muon momentum selection and to steer the beam into the MICE experimental hall~\cite{beamline}. The DS increases the muon rate by keeping the pions in the solenoid until they decay. As a result, the beam that enters MICE has a small pion content. To ensure further muon beam purity, these pions are tagged and rejected from the cooling measurement using a series of particle identification (PID) detectors (Fig.~\ref{fig:MICE_Step4})~\cite{pion}. 
\section{MICE Detectors}
\begin{figure}[tbh]
    \centering
    \includegraphics[width=1\columnwidth]{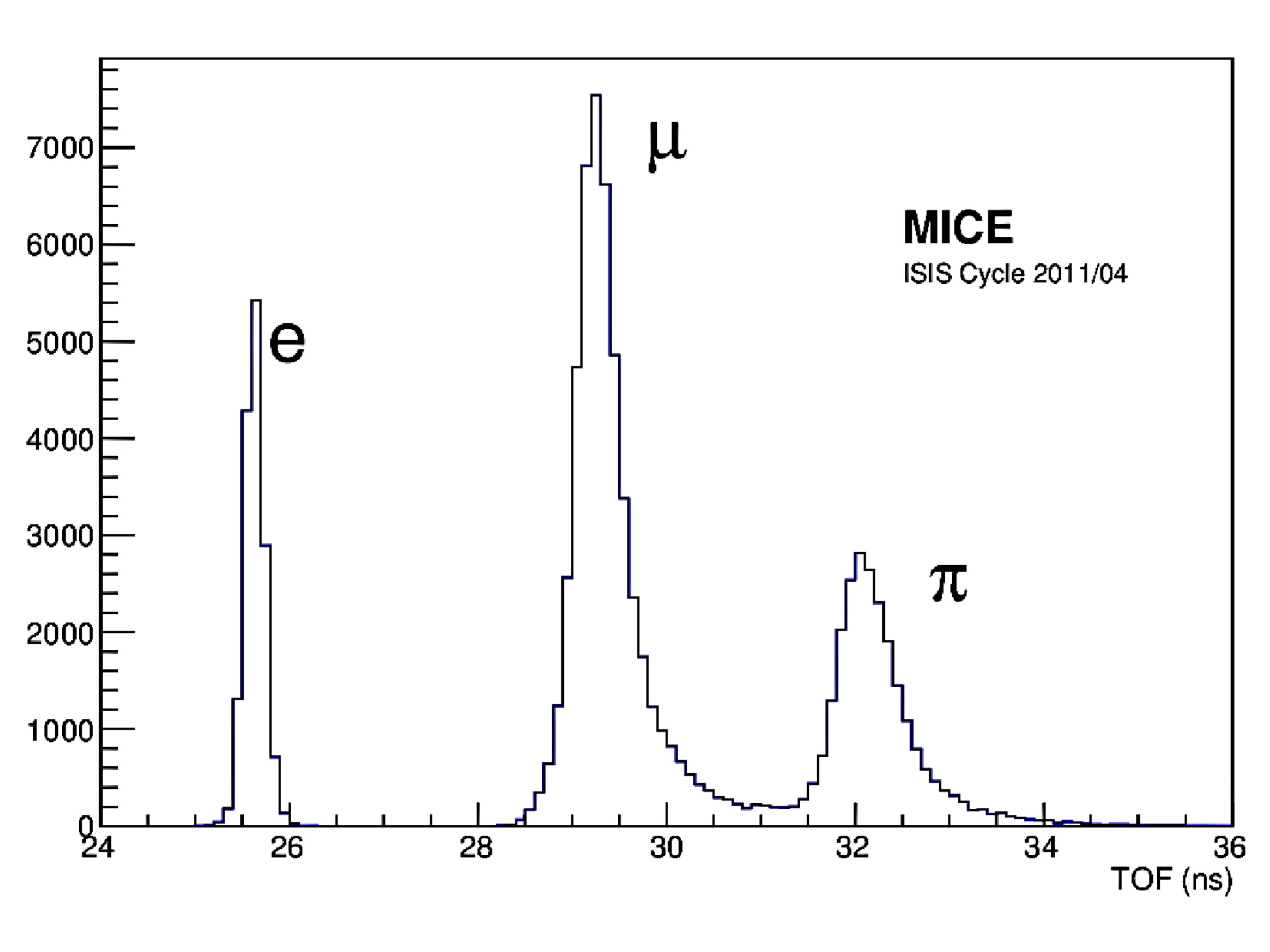}
    \caption{The measured electron, muon, and pion time-of-flight distributions between the ToF0 and ToF1 detectors~\cite{pion}.}
   \label{TOF_Data}
\end{figure}

\begin{figure}[tbh]
    \centering
    \includegraphics[width=1\columnwidth]{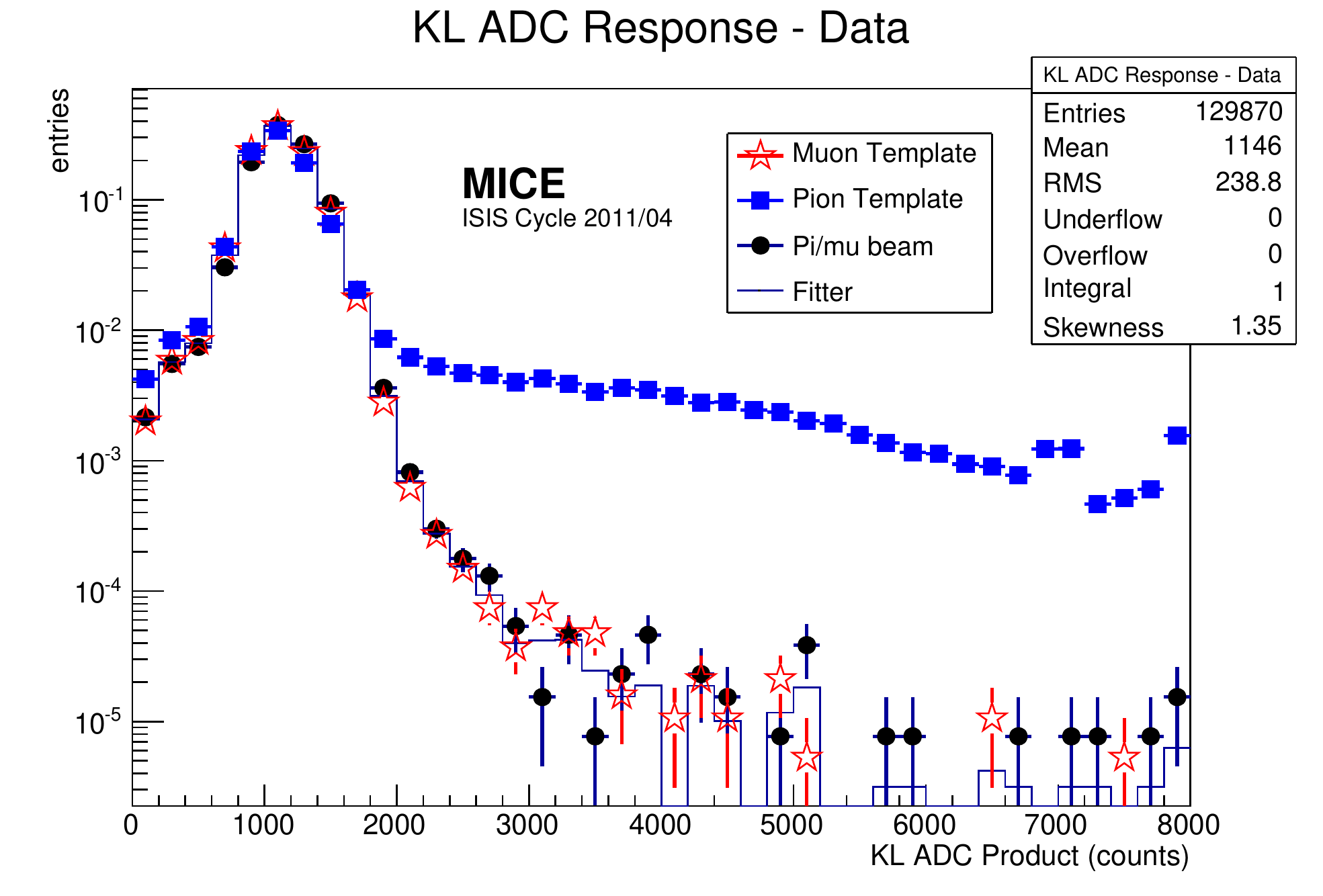}
    \caption{Histogram of the KLOE-Light detector ADC count~\cite{pion}.}
    \label{KL_Data}
\end{figure}

\begin{figure}[tbh]
    \centering
    \includegraphics[width=1\columnwidth]{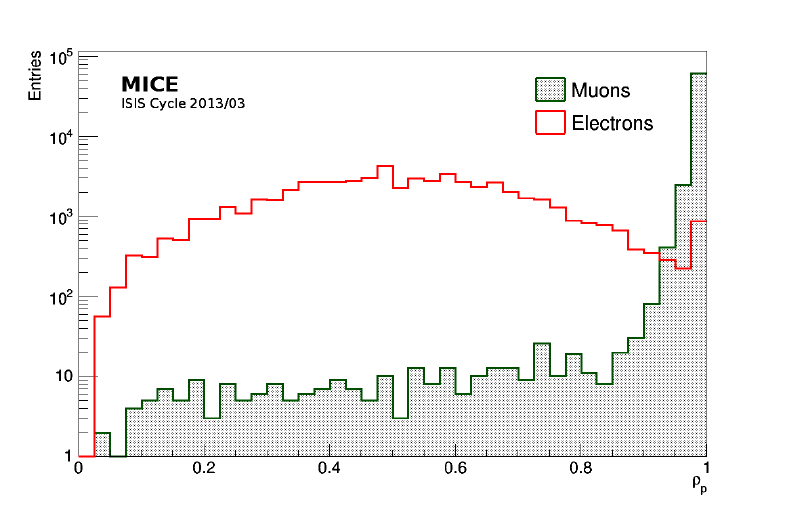}
    \caption{Histogram of the muon and electron plane densities as measured with the EMR calorimeter~\cite{EMR}.}
    \label{EMR_Data}
\end{figure}

MICE detectors can be divided into PID or tracking detectors. The primary MICE PID detectors are three Time of Flight (ToF) counters (ToF0, ToF1, and ToF2) made of scintillating slabs. Electrons, with smaller masses, and pions, with larger masses, have respectively shorter and longer times of flight than muons and, as a result, they each have a distinct time-of-flight peak (Fig.~\ref{TOF_Data})~\cite{pion}. In addition, two aerogel Cherenkov detectors with differing refractive indices are used to reject pions and electrons. The KLOE-Light (KL) detector is a calorimeter and uses lead layers alternating with scintillating fibers to differentiate between particles based on their energy depositions. Electrons shower in KL and as a result have broader KL distributions compared to muons and pions (Fig.~\ref{KL_Data}). The most downstream calorimeter is the Electron Muon Ranger (EMR), made of planes of scintillator extrusions, makes use of the differing particle track topologies. Electron showers in the EMR typically miss some EMR planes as they travel along their paths (occupancy < $1$) in contrast to muon tracks which consistently hit all planes (occupancy $\simeq  1$) (Fig.~\ref{EMR_Data})~\cite{EMR}. 

MICE has two tracking detectors, upstream and downstream of the absorber~\cite{tracker}. Each tracker is composed of five planar scintillating-fiber stations, each with three doublet fiber layers, and is immersed in the solenoidal field of a Spectrometer Solenoid (SS). The upstream and downstream SS consist of five superconducting coils; two match the muon beam into and out of the absorber (Match 1 and Match 2 coils), and three produce constant fields inside the tracking volumes (End 1, Center, and End 2 coils). For measurement of beam cooling in MICE, the input and output beam distributions are compared at the tracker stations immediately upstream and downstream of the absorber (tracker reference planes). 
\section{Measurement of Muon Ionization Cooling with MICE}
Ionization cooling occurs when muons travel through an absorbing material and undergo a ``soft'' inelastic collision. With the incoming muon energies significantly larger than the binding energies of the atomic electrons of the material, the atom ionizes, ejecting a valence electron~\cite{energy_loss}. Concurrent with the ionization energy loss process, elastic collisions with the atomic nuclei of the material can occur. Given the larger mass of the atomic nuclei compared with the incident muons, this process causes very small energy loss. However, it is the driving force behind the scattering of muons by small angles~\cite{energy_loss}. This process is known as multiple Coulomb scattering and is a source of beam heating in MICE. \raggedbottom

A good understanding of energy loss and multiple scattering is essential in ionization cooling where the muon momentum is reduced in all directions (both transversely and longitudinally) with a subsequent restoration of the beam longitudinal momentum in RF cavities, suitable for a multi-segment cooling system. The beam cooling equation that describes this process is written in terms of the rate of change of normalized transverse root-mean-square (RMS) emittance,~$\varepsilon_{\perp}$~\cite{dave}:
\begin{equation}\label{eq:cooling}
\frac{d\varepsilon_{\perp}}{dx} \approx -\frac{\varepsilon_{\perp} }{\beta ^{2}E_{\mu }}\left \langle \frac{dE}{dx} \right \rangle +\frac{\beta _{\perp} (13.6 \ \textnormal{MeV}/c)^{2}} {2\beta ^{3}E_{\mu }m_{\mu }X_{0}},
\end{equation}
where $E_{\mu}$ is the muon energy, $\beta c$ the muon velocity, $dE/dx$ the magnitude of the ionization energy loss, $m_{\mu}$ the muon mass, $X_{0}$ the radiation length, and $\beta_{\perp}$ the transverse beta function at the absorber. 

The first term in the equation represents cooling from ionization energy loss and the second term describes heating from multiple scattering. The minimum achievable emittance, or equilibrium emittance, for a given material and focusing conditions is obtained by setting the rate of change of the normalized transverse RMS emittance to zero:
\begin{equation}\label{eq:equilibrium}
\varepsilon_{\textnormal{equilibrium}} \cong \frac{\beta_{\perp }(13.6 \ \textnormal{MeV}/c)^{2} }{2X_{0} \left \langle \frac{dE}{dx} \right \rangle \beta m_{\mu }}.
\end{equation}

A smaller equilibrium emittance leads to a more effective emittance reduction, which from Eq.~\ref{eq:equilibrium} occurs when $\beta_{\perp}$ is minimized and $X_{0}\left \langle \frac{dE}{dx} \right \rangle$ maximized~\cite{dave}. In MICE, a small beta function is achieved by focusing the beam tightly at the absorber, and large $X_{0}\left \langle \frac{dE}{dx} \right \rangle$ is achieved by the use of low-$Z$ (low atomic number) absorbing materials such as lithium hydride (LiH) and liquid hydrogen (LH$_{2}$). 
\section{Muon Ionization Cooling Results}
\begin{figure}[tbh]
    \centering
    \includegraphics[width=1\columnwidth]{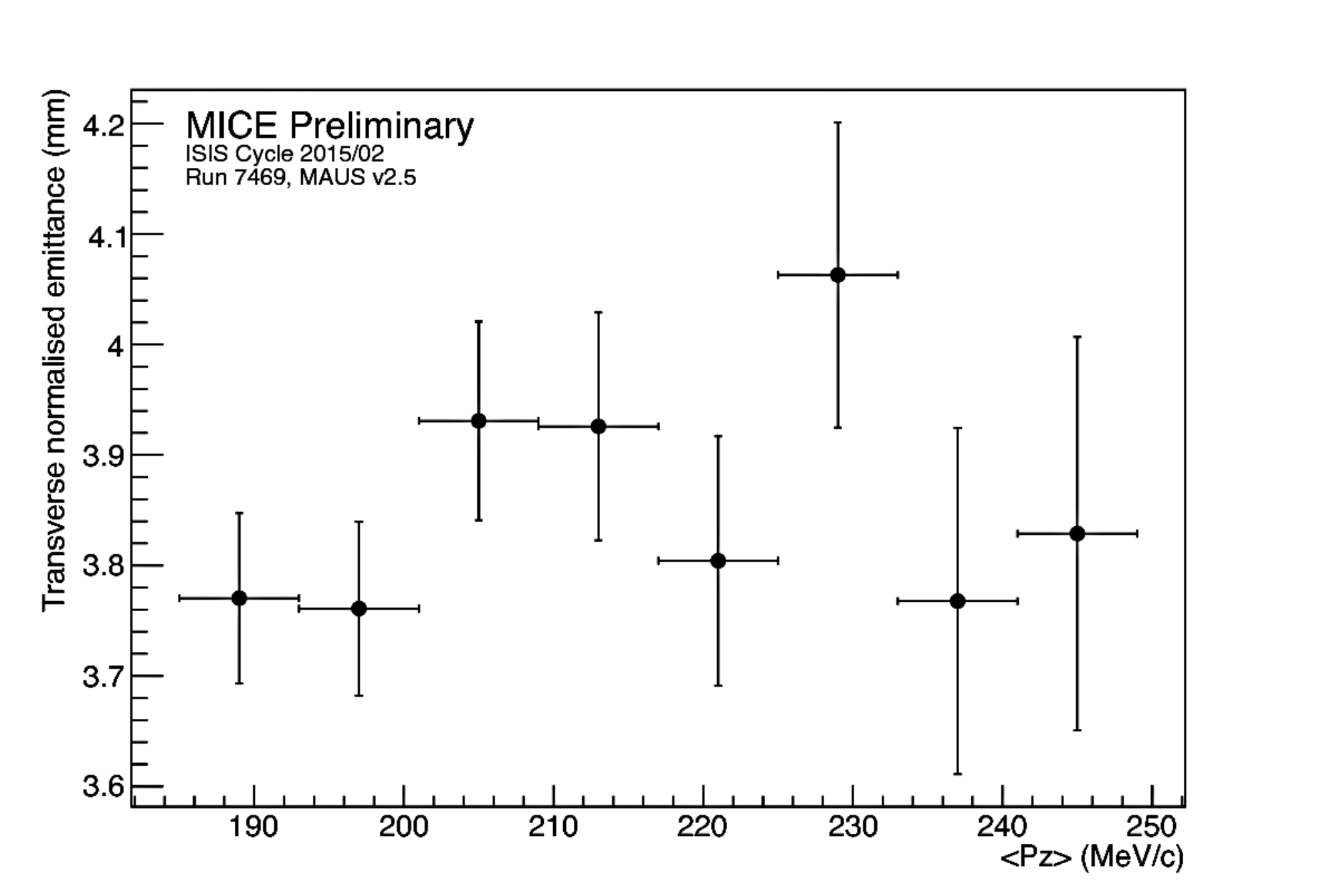}
    \caption{First direct measurement of emittance using the most upstream scintillating fiber tracker station~\cite{emittance}.}
    \label{fig:first_emittance}
\end{figure}
\begin{figure}[tbh]
    \centering
    \includegraphics[width=1\columnwidth]{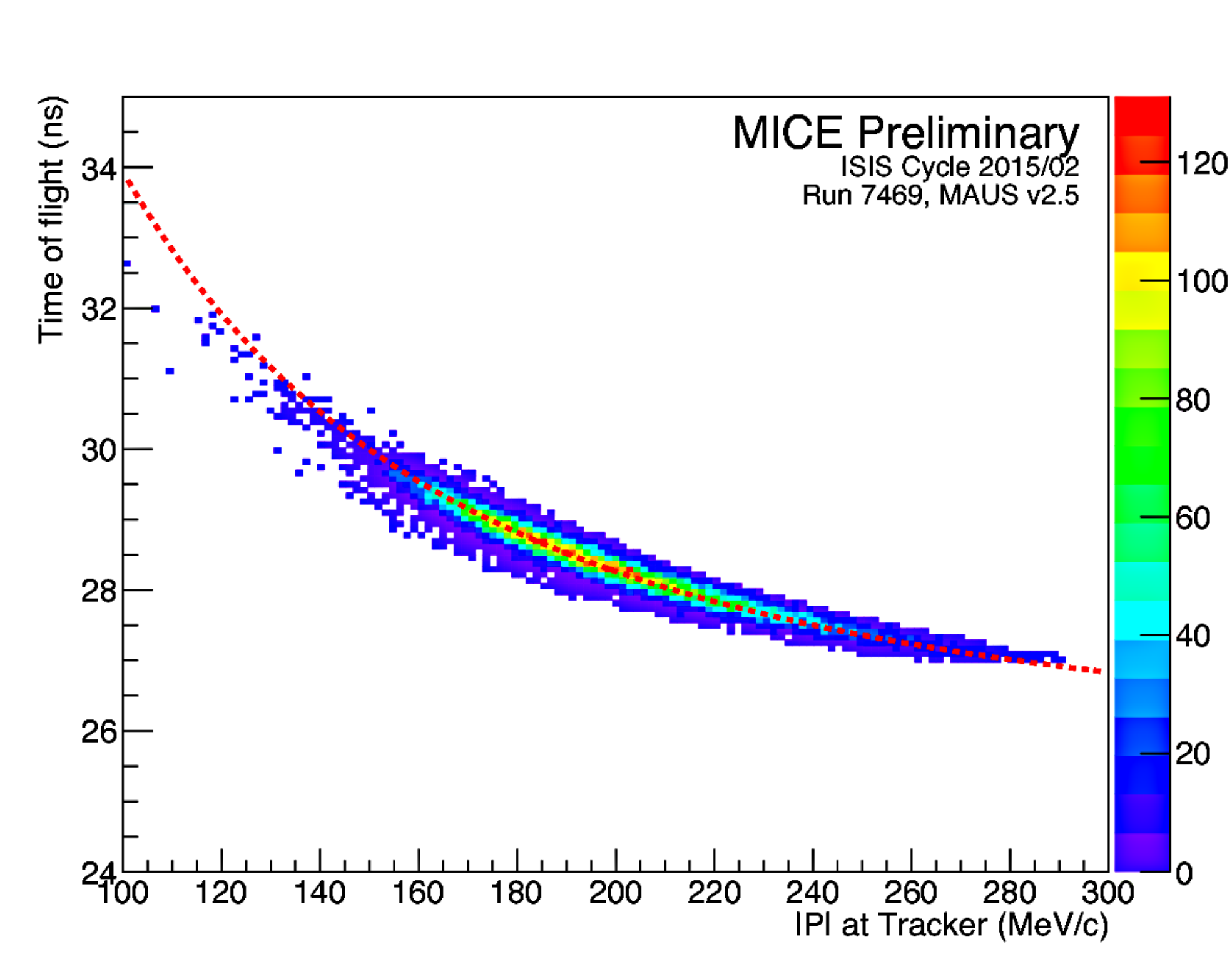}
    \caption{Comparison of momentum and time of flight between TOF0 and TOF1 for the particle sample used to make the first direct measurement of emittance. The red curve represents the
muons that have the mean momentum loss between
ToF1 and the tracker~\cite{emittance}. }
    \label{fig:beam_selection}
\end{figure}
\begin{figure*}[tbh]
    \centering
    \includegraphics*[width=0.8\textwidth]{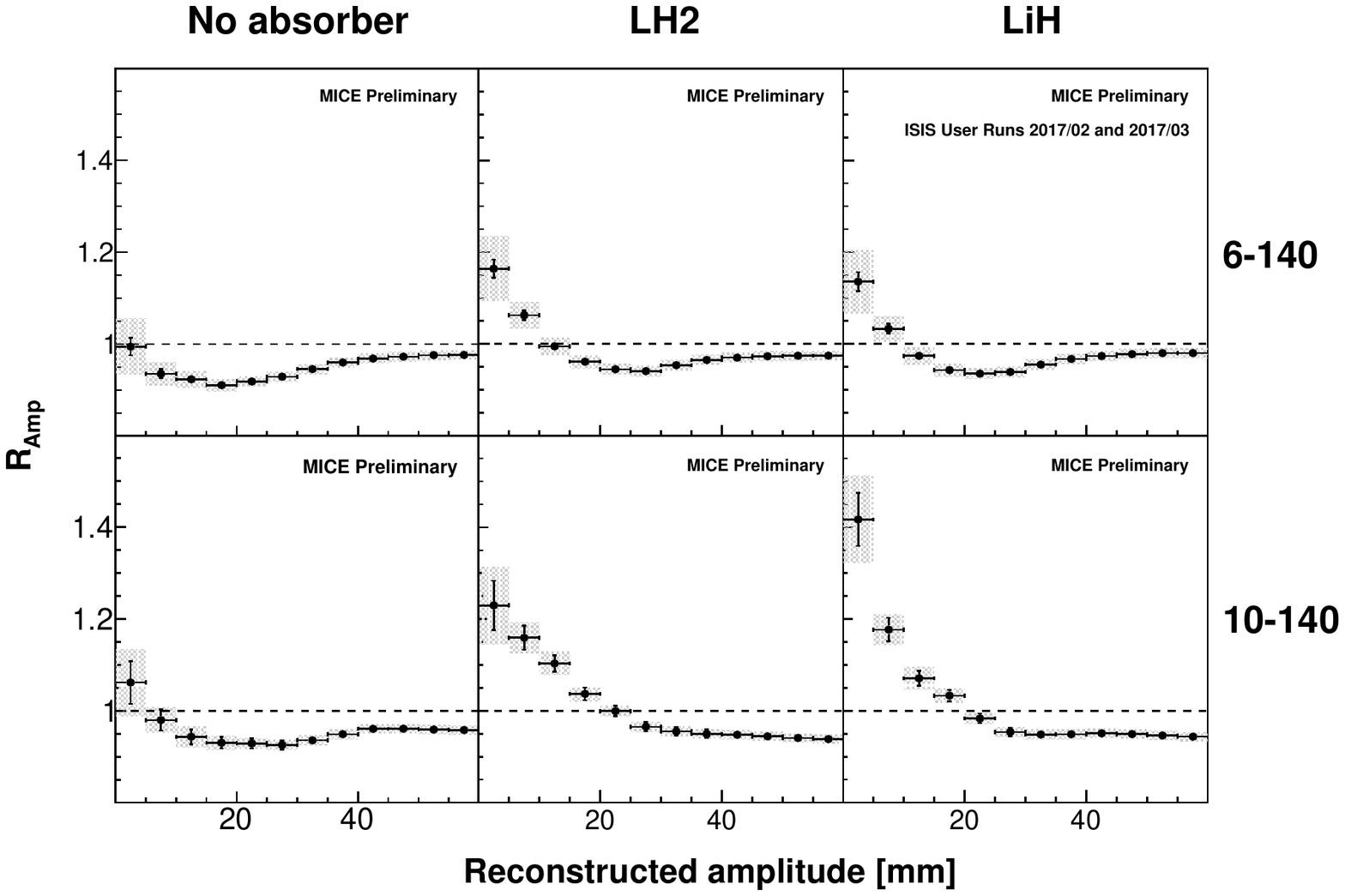}
    \caption{Ratio of cumulative amplitude distributions between upstream and downstream tracker reference plane for the 6-140 (top) and the 10-140 (bottom) nominal beam settings, and three absorber configurations of ``no absorber'' (empty channel, left), LH$_{2}$ (middle), and LiH (right).}
    \label{fig:amplitude}
\end{figure*}
One figure of merit for muon beam cooling in MICE is the reduction of the normalized transverse RMS emittance. MICE is a single particle experiment and the phase-space coordinates in $(x, p_{x}, y, p_{y})$ of each muon are reconstructed using the trackers. From the measured phase-space coordinates, the normalized transverse RMS emittance can be calculated, 
\begin{equation}
\varepsilon_{\perp}=\frac{\left | \Sigma \right |^{\frac{1}{4}}}{m_{\mu}},
\end{equation}
where $\Sigma$ is the covariance matrix,
\begin{equation}\label{eq:covariance}
\Sigma=\left(
\begin{matrix}
\sigma_{xx} & \sigma_{p_{x}x} & \sigma_{yx} & \sigma_{p_{y}x} \\ 
\sigma_{xp_{x}} & \sigma_{p_{x}p_{x}} & \sigma_{yp_{x}} & \sigma_{p_{y}p_{x}} \\ 
\sigma_{xy} & \sigma_{p_{x}y} & \sigma_{yy} & \sigma_{p_{y}y} \\ 
\sigma_{xp_{y}} & \sigma_{p_{x}p_{y}} & \sigma_{yp_{y}} & \sigma_{p_{y}p_{y}}
\end{matrix}\right).
\end{equation}

Figure~\ref{fig:first_emittance} displays the muon beam emittance versus momentum. It is the first measurement of emittance using the MICE upstream tracker. The emittance is observed to be approximately constant within the momentum range $180$ MeV/$c$ to $250$ MeV/$c$. The MICE PID detectors were used to select a pure beam of muons. This is demonstrated in Fig.~\ref{fig:beam_selection} where the time-of-flight from ToF0 to ToF1 is plotted versus the reconstructed momentum in the upstream tracker. The selected particles display the correlation expected for a beam composed of muons. 

In addition to emittance reduction, the MICE single particle measurement capability allows the development of alternative beam cooling figures of merit. One such figure of merit is reduction of the single-particle amplitude, defined as the weighted distance of each muon from the beam center. Single-particle amplitude can be used to probe the change in density in the core of the beam, where an increase in phase-space density signifies beam cooling. The transverse beam amplitude can be obtained using the measured transverse emittance~\cite{chris}, 
\begin{equation}\label{eq:amplitude}
A_{\perp}=\varepsilon_{\perp}\left(\vec{v}-\vec{u}\right)^{T} \Sigma^{-1}\left(\vec{v}-\vec{u}\right),
\end{equation}
where $\varepsilon_\perp$ is the normalized transverse RMS emittance, $\Sigma$ is the covariance matrix (Eq.~\ref{eq:covariance}), $\vec{v}$ is the transverse phase-space vector ($x$, $p_x$, $y$, $p_y$), and $\vec{u}$ is the beam centroid.  

\begin{figure}[tbh]
    \centering
    \includegraphics[width=1\columnwidth]{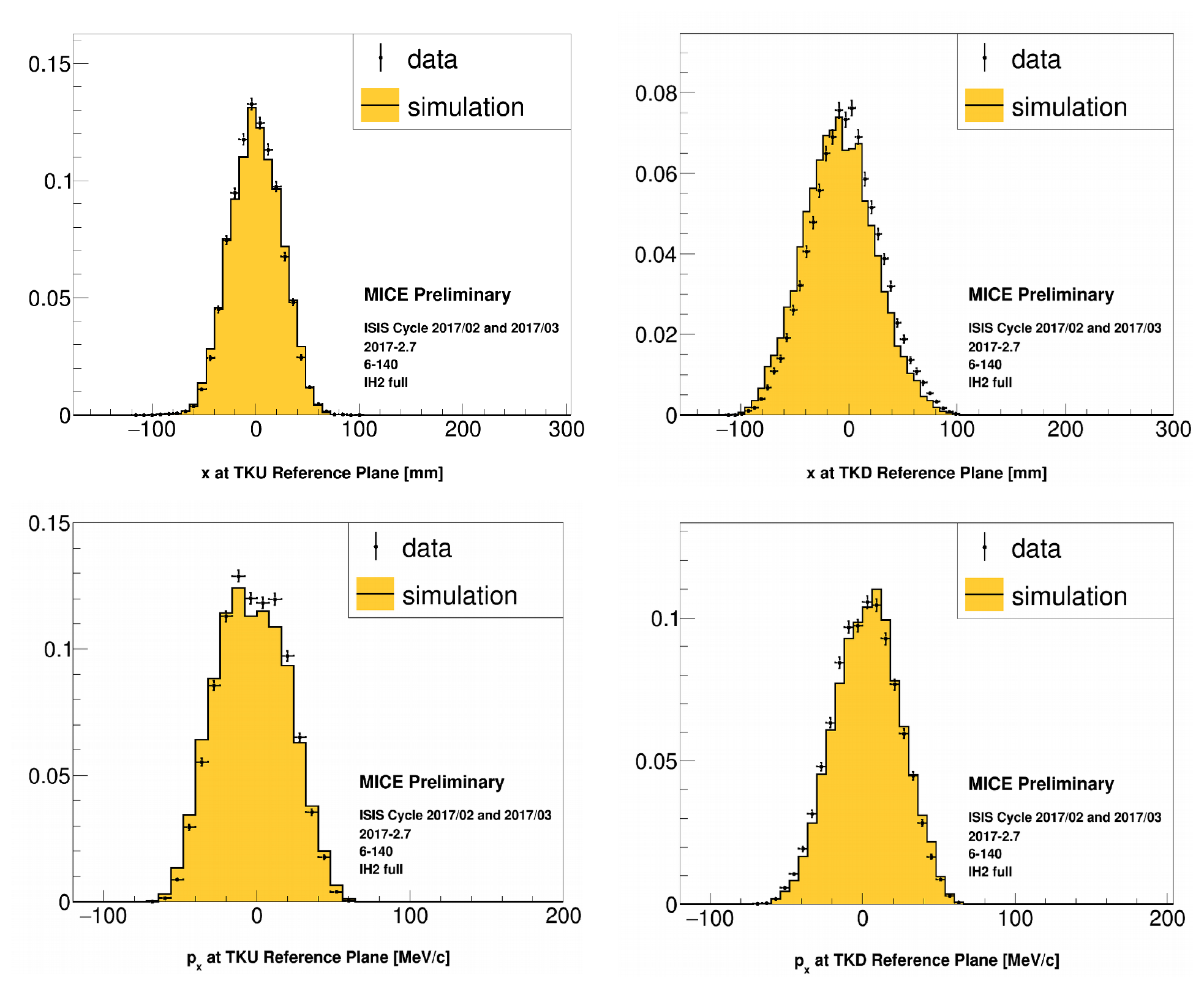}
    \caption{Comparison of $x$ (top) and $p_{x}$ (bottom) distributions at the upstream (left) and downstream (right) tracker reference planes in data and simulation (6-140, LH$_{2}$ absorber setting).}
    \label{fig:systematics_xpx}
\end{figure}

\begin{figure}[tbh]
    \centering
    \includegraphics[width=1\columnwidth]{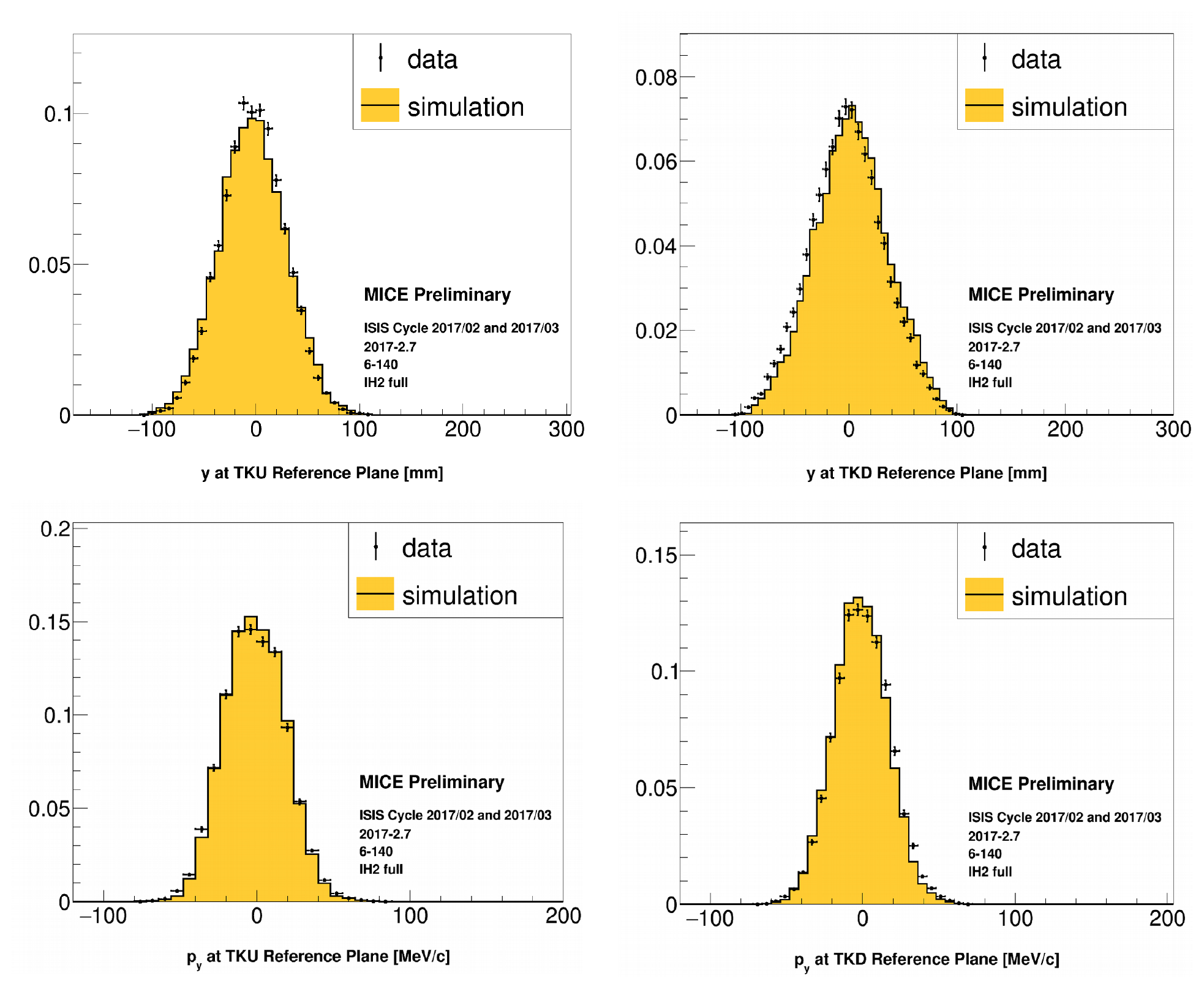}
    \caption{Comparison of $y$ (top) and $p_{y}$ (bottom) distributions at the upstream (left) and downstream (right) tracker reference planes in data and simulation (6-140, LH$_{2}$ absorber setting).}
    \label{fig:systematics_ypy}
\end{figure}

The ratios of the cumulative amplitude distributions are calculated from the measured coordinates and momenta at the upstream and downstream tracker reference planes (Fig.~\ref{fig:amplitude}) for initial normalized transverse RMS emittances of $6$ and $10$ mm and reference momentum of $140$ MeV/$c$ (referred to as 6-140 and 10-140 settings). The optics of the runs used in this study had a transverse beta function, $\beta_{\perp}$ of $660$ mm at the absorber and their corresponding $x$, $p_{x}$, $y$, and $p_{y}$ distributions are shown in Figs.~\ref{fig:systematics_xpx} and~\ref{fig:systematics_ypy} with a comparison with simulation~\cite{amplitude}. The amplitude change has been studied for two absorber materials, LH$_{2}$ and LiH, as well as an empty channel setting (``no absorber'' in Fig.~\ref{fig:amplitude}). The empty channel measurement enables a cross-check with the Liouville theorem where the amplitude measurement is expected to stay approximately constant. The ratio of cumulative amplitude distributions ($R_{Amp}$) in this context is $\frac{n_{DS}}{n_{US}}$ where $n$ is the number of muons with amplitude equal to or less than the corresponding amplitude on the horizontal axis. The muon count within each amplitude bin represents the density associated with the bin and an increase in the density at the beam core (low amplitude region of the beam) from upstream to downstream tracker reference planes (due to migration of high amplitude muons to low amplitude) indicates beam cooling, i.e. as $R_{Amp} > 1$ in Fig.~\ref{fig:amplitude}. 
\section{Conclusion}
We have presented the first measurement of ionization cooling in MICE, an important R\&D step towards a future muon collider, neutrino factory, and other cooled muon applications.~The measurements of the ratio of the upstream and downstream cumulative amplitude distributions demonstrate cooling (migration of large transverse amplitude muons to lower amplitudes after crossing the absorber). MICE has concluded all data taking and is currently in the analysis phase, with several publications in preparation. 
\section{Acknowledgement}
MICE has been made possible by grants from DOE, NSF (U.S.A), the INFN (Italy), the STFC (U.K.), the European Community under the European Commission Framework Programme 7 (AIDA project, grant agreement no.~262025, TIARA project, grant agreement no.~261905, and EuCARD), the Japan Society for the Promotion of Science and the Swiss National Science Foundation, in the framework of the SCOPES programme. We acknowledge the support of the staff of the STFC Rutherford Appleton and Daresbury Laboratories, and the use of Grid computing resources deployed and operated by GridPP in the U.K., http://www.gridpp.ac.uk/. The author acknowledges the support that she has received from United States National Science Foundation, the Division of Physics of Beams of the American Physical Society, and TRIUMF to attend IPAC 2018. 
\atColsBreak{

\end{document}